\documentstyle[revtex]{aps}
\begin{document}
\draft
\begin{title}
Localization of shadow poles by complex scaling
\end{title}
\author{Attila Cs\'ot\'o\cite{email}}
\begin{instit}
Institute of Nuclear Research of the Hungarian Academy of
Sciences \\ P.O.Box 51 Debrecen, H--4001, Hungary
\end{instit}
\receipt{5 May 1993}
\begin{abstract}
Through numerical examples we show that the complex scaling
method is suited to explore the pole structure in
multichannel scattering problems. All poles lying
on the multisheeted Riemann energy surface, including shadow
poles, can be revealed and the Riemann sheets on which they
reside can be identified.
\end{abstract}


Resonant states are solutions of the Schr\"odinger equation with
outgoing asymptotic boundary condition. It was
pointed out long ago that these solutions must belong to
complex eigenenergies, and the scattering matrix has poles at
these energies \cite{Siegert}. In the coordinate space resonant
eigenfunctions show oscillatory behavior in the asymptotic
region, with exponentially growing amplitude, $\sim\exp[i(\kappa
-i\gamma)r]$\ $(\kappa ,\gamma >0)$, thus they are not elements
of the $L^2$ space. Complex scaling is a most powerful and
easily applicable method to
describe such states \cite{Ho,Moiseyev,Reinhard}. It has been
successfully applied in atomic \cite{Ho,Resonances} and nuclear
physics \cite{Resonances,Kruppabe8,KruppaKato}.

In single-channel problems the working mechanism of the complex
scaling method (CSM) is well understood and there is almost no
obscure point. The situation is not so clear, however, in
multichannel cases. The pole structure of a multichannel
scattering matrix is much more complicated than that of a
single-channel $S$-function. In the case of Hermitean potentials
a pole that would appear in
one of the $N$ channels in a single-channel problem, gives rise
to $2^{N-1}$ poles on different Riemann energy sheets
in the coupled $N$-channel problem \cite{Eden,Pearce}. The
easiest way to label a Riemann sheet is to give the signs of the
imaginary parts of the channel wave numbers $k_i$
$(i=1,2,\dots,N)$ in an $N$-term sign-string (sign(Im
$k_1$),$\;$sign(Im $k_2$),...,sign(Im $k_N$)) \cite{CsotoPRL}.
In the zero-coupling limit all $2^{N-1}$ poles are at the same
energy position on different sheets, while, by varying the
coupling strengths, the
poles move, and a crossing of the real energy axis by one of
them implies a crossing over to another Riemann sheet.
It has been a long-standing belief that only those poles
(named ordinary poles) can have appreciable effects on the
physically observable quantities which are on the Riemann sheet
adjacent to the physical one. Recently, the effect of other
poles, the so-called shadow poles, of the multichannel scattering
matrix on some physical observables has attracted interest in
atomic \cite{Shakesaft}, particle \cite{Morgan}, and nuclear
physics \cite{CsotoPRL,Hale}. It turned out that in certain
cases the shadow poles can cause strong effects. E.g.,\ it is a
shadow pole which causes the very large cross section of the
famous $d+t$$\rightarrow$$\alpha +n$ thermonuclear reaction
\cite{CsotoPRL,Hale}.
The effect of a shadow pole on the scattering matrix depends
crucially on which Riemann sheet it is situated
\cite{Pearce,CsotoPRL,Hale}.

Although most of the applications of the CSM are in multichannel
problems, up till now all investigations have been concerned
with ordinary
poles. In this Brief Report we show that, using the CSM, one can
search for poles on different Riemann sheets and can identify
the poles by their sheets.

In a one-channel case the essence of the CSM is as follows.
Instead of the
\begin{equation}
\widehat H\vert\Psi\rangle=(\widehat T+\widehat
V)\vert\Psi\rangle =E\vert\Psi\rangle
\label{Sch}
\end{equation}
eigenequation of the Hamiltonian $\widehat H$, we solve the
eigenvalue problem of the transformed Hamiltonian $\widehat
H_\theta=\widehat U(\theta)\widehat H\widehat U^{-1}(\theta)$:
\begin{equation}
\widehat H_\theta\vert\Psi_\theta\rangle=E_\theta\vert
\Psi_\theta\rangle
\label{Scht}
\end{equation}
(the $\theta$ subscript of $\Psi$ means that
the wave function implicitly depends on $\theta$; Hamiltonians
with different $\theta$ results in different wave functions).
$\widehat U(\theta)$ is an unbounded similarity
transformation \cite{Lowdin}, which, in
the coordinate space, acts on a function $f(r)$ such that
\begin{equation}
\widehat U(\theta)f(r)=e^{3i\theta/2}f(re^{i\theta}).
\label{CS}
\end{equation}
(If $\theta$ is real, $U(\theta)$ means a rotation into the
complex coordinate plane, if it is complex, it means a rotation
and scaling.) The two problems are connected by the ABC theorem
\cite{ABC}. If $\widehat V$ is a (dilation) analytic operator:
\begin{enumerate}
 \item[(i)] The bound eigenstates of $\widehat H$ are the
eigenstates of $\widehat H_\theta$, regardless of the actual
value of $\theta$, within $0\leq\theta<\pi/2$;
\item[(ii)] the continuous spectrum of $\widehat H$ will be
rotated by an angle 2$\theta$;
 \item[(iii)] a complex generalized eigenvalue of
Eq.\ (\ref{Scht}), $E_{\rm res}=\varepsilon
-i{{1}\over{2}}\Gamma$, $\varepsilon ,\Gamma >0$ (with the wave
number $k_{\rm res}=\kappa -i\gamma,$ $\kappa ,\gamma >0$),
belongs to the proper spectrum of $\widehat H_\theta$ provided
$2\theta >\vert \arg E_{\rm res}\vert$.
\end{enumerate}
Roughly speaking, the complex scaling transformation changes the
asymptotic wave function from $\exp[i(\kappa -i\gamma)r]$ to
$\exp[i(\kappa -i\gamma)r\exp(i\theta)]$, which, in the case
of $2\theta >\vert \arg E_{\rm res}\vert =2\vert
\arg k_{\rm res}\vert$, makes the diverging wave function
localized. It is important to note that, if the sign of $k_{\rm
res}$ is reversed, then the outgoing wave with $\theta =0$ is
localized, and the complex scaling spoils the localization
unless $2\theta <\vert \arg E_{\rm res}\vert$. In a
single-channel problem with a Hermitean potential $-k_{\rm res}$
is on the physical sheet, where there are no resonance poles,
but in a multichannel problem the imaginary parts of some
channel wave numbers may be positive.

In multichannel cases Eq.\ (\ref{Sch}) becomes a matrix equation
\begin{equation}
\sum_{\beta=1}^N\widehat H_{\alpha\beta}\vert\Psi^\beta\rangle =
E_\alpha\vert\Psi^\alpha\rangle ,\ \ \ \alpha =1,2,\dots,N,
\label{MSch}
\end{equation}
where the Greek letters are the channel indices. The
transformation operator of Eq.\ (\ref{CS}) becomes
\begin{equation}
\widehat U_{\alpha\beta}(\theta)=\delta_{\alpha\beta}\widehat U
(\theta).
\label{MCSM}
\end{equation}

In the literature there are some hints on the strange
behavior of the CSM in multichannel cases. E.g.\ in
\cite{Elander} the authors found that varying the rotation angle
$\theta$, channel continua can absorb resonances that were
revealed before, however, they did not explain this phenomenon.
In \cite{Reinhard} it is stated that such a phenomenon becomes
transparent if one studies the multichannel problem on the
Riemann energy surface, but no attempt has been made to assign
these poles to Riemann sheets.

Here we study the working mechanism of the multichannel CSM in a
simple model, which is easy to comprehend and control.
Our model consists
of a target with two internal states, whose thresholds are $E_1$
and $E_2$, respectively, and a projectile. The two target states
are the two channels. We choose one-term separable potentials
\cite{CsotoPR} for both the diagonal and interchannel
interactions,
\begin{equation}
\widehat V_{\alpha\beta}=\vert \varphi^\alpha_0(b)\rangle
\lambda_{\alpha\beta}\langle \varphi^\beta_0(b) \vert ,\ \ \
\alpha,\beta =1,2,
\label{pot}
\end{equation}
where $\vert\varphi_0(b)\rangle$ is the eigenfunction of the
three dimensional harmonic oscillator with $n=0$ oscillator
quantum, $b$ is the size parameter and $\lambda _{\alpha\beta}$
are the (real) potential strengths
($\lambda_{12}=\lambda_{21}$). For the wave functions
$\vert\Psi^\gamma_\theta\rangle$ of (\ref{Scht}), we use the
following trial functions
\begin{equation}
\vert\Psi^\gamma_\theta\rangle =
\sum_{i=0}^{n_\gamma}c_{\gamma i}\vert\varphi^\gamma_i(\bar
b)\rangle, \ \ \gamma =1,2
\label{WFE}
\end{equation}
in a variational method for the expansion coefficients $c_i$
(this is the well-known wave function expansion (WFE) method).
With this ansatz, functions $\vert\Psi^\gamma_\theta\rangle\in
L^2$ are selected.
The use of harmonic oscillator functions both in (\ref{pot}) and
(\ref{WFE}) makes it possible to calculate all necessary matrix
elements analytically \cite{CsotoPR}. We choose size
parameters in (\ref{WFE}) different from that in (\ref{pot}) so
as to make the trial function more flexible. If we set the
strength of one of
the diagonal interactions to be zero, the selected channel
cannot accommodate a resonance, so that all poles we find in the
coupled-channel problem must originate from the other channel,
which implies that the sign belonging to the other channel must
be negative. For the sake of simplicity, we take
$s$-wave states throughout; although, the analytical expressions
of \cite{CsotoPR} can be used for $l\ne 0$ as well.

As a first example, we choose $\lambda_{22}=0$,
$\lambda_{11}=1.0$, $b=0.6$, $\bar b=2.0$, $E_1=0$, and
$E_2=2$ (we use atomic units $\hbar =m=1$). The basis sizes
($n_1$, $n_2$) are chosen so
as to reach stable convergence. In the uncoupled case
($\lambda_{12}=\lambda_{21}=0$) only one resonance pole appears
at $E=3.049-i2.153$ [Fig.\ \ref{fig1}(a)]. Switching on
the coupling ($\lambda_{12}=\lambda_{21}=1.0$), we get different
pole arrangements
at different rotation angles [Figs.\ \ref{fig1}(b)--(d)]. We can
see in all figures that the poles distort the continua, as if
they attracted or repelled the continuum points. This phenomenon
has surfaced several times earlier, e.g.\ Refs.\
\cite{Kruppabe8,CsotoFB}, but as far as the author knows, it is
yet unexplained. Furthermore, we emphasize
that in this work the central question is the working
mechanism of the CSM in multichannel problems, which requires
the use of very different rotation angles. Thus we do
not perform an optimization in the $\theta$ parameter (which
could be done by choosing the stationary point of the
$\theta$--trajectory, see e.g.\ \cite{Kruppabe8}). These figures
show the appearance and disappearance of poles, the same
phenomenon
as was mentioned above. The choice $\lambda_{22}=0$ and the fact
that in the uncoupled case there is only one resonance pole
guarantees that in this problem there is a pole on the $(--)$
Riemann sheet and another, a shadow pole, on $(-+)$
\cite{Pearce}. The pole at $4.742-i1.810$ is revealed only when
both continua have swept over this point, which
implies that the condition $2\theta >\vert \arg E^\gamma_{\rm
res}\vert$, $\gamma =1,2$ must be fulfilled in both channels
(where $E^1_{\rm res}$ and $E^2_{\rm res}$ are the channel
energies). Consequently, this pole is on the $(--)$ sheet. For
the other pole, which is on the $(-+)$ sheet at $2.395-i1.467$,
the relations $2\theta >\vert \arg E^1_{\rm res}\vert$ and
$2\theta <\vert \arg E^2_{\rm res}\vert$ must hold (cf.\ the
remark after the ABC theorem, above). This is in full agreement
with what we can see in the figures.

This example tells us that, if a pole is revealed, then it is
on a Riemann sheet which is characterized by negative signs for
all channels whose continua have swept over the pole, and
positive signs for all others. A pole at $E$ is an
ordinary one (i.e.\ it is on a sheet adjacent to the
physical sheet) if it has been swept over by the continua of all
channels whose threshold energies are lower than Re$(E)$ and has
not been swept over by any other ones. From this it follows that
one can imagine situations where a shadow pole can be revealed
only if the rotation angles in different channels are different.
For instance, to reveale a pole above the first channel
threshold on the $(+-)$ sheet, the rotation angle in
the second channel must be greater than the one in the first
channel. It is questionable whether the CSM can cope with such a
constraint. The theory of the multichannel CSM always assumes
that $\theta$ is the same for all channels. Here, just to see
what happens, we venture to choose two different $\theta$.

It seems natural to generalize the multichannel complex scaling
transformation (\ref{MCSM}) in the following way:
\begin{equation}
\widehat U_{\alpha\beta}=\delta_{\alpha\beta}\widehat
U_\alpha(\theta_\alpha).
\label{MCHU}
\end{equation}
In the coordinate space the action of $\widehat
U_\alpha(\theta_\alpha)$ is
\begin{equation}
\widehat U_\alpha(\theta_\alpha)f(r)=
e^{3i\theta_\alpha/2}f(re^{i\theta_\alpha}).
\end{equation}
This definition ensures that the $\{\widehat U_{\alpha\beta} \}$
operator matrix inherits all
properties of $\widehat U(\theta)$. Applying the transformation
(\ref{MCHU}) to the Hamiltonian of Eq.\ (\ref{MSch}), we arrive
at
\begin{eqnarray}
\sum_{\beta =1}^N\widehat U_{\alpha}(\theta_\alpha)\widehat
H_{\alpha\beta} \widehat
U_{\beta}^{-1}(\theta_\beta)\vert\Psi^\beta_{\theta_\beta}\rangle
&=&\nonumber\\
E_{\theta_\alpha,\theta_\beta}^\alpha\vert
\Psi^\alpha_{\theta_\alpha} \rangle ,\ \ \alpha
&=&1,2,\dots,N.
\end{eqnarray}
As an illustrative example, we write down the function $\widehat
U_{\alpha}(\theta_\alpha)\widehat H_{\alpha\beta} \widehat
U_{\beta}^{-1}(\theta_\beta)\vert\Psi^\beta_{\theta_\beta}\rangle$
and its overlap with the function
$\langle\Psi^\alpha_{\theta_\alpha}\vert$ in the
coordinate space. If the operator $\widehat H_{\alpha\beta}$
connects channels which have the same dynamical coordinate $r$,
then
\begin{eqnarray}
\langle r\vert\widehat U_{\alpha}(\theta_\alpha)\widehat
H_{\alpha\beta} \widehat
U_{\beta}^{-1}(\theta_\beta)\vert\Psi^\beta_{\theta_\beta}\rangle
=\ \ \ \ \ \ \ \ \ \ \ \ \ \ \ \ \ \ \ \ \ \ \ &&\nonumber\\
e^{3i(\theta_\alpha-\theta_\beta)/2}
H_{\alpha\beta} (re^{i\theta_\alpha})\Psi^\beta_{\theta_\beta}
(re^{i(\theta_\alpha-\theta_\beta)}), \ \ \ \ &&
\end{eqnarray}
and
\begin{eqnarray}
\langle \Psi^\alpha_{\theta_\alpha}\vert\widehat
U_{\alpha}(\theta_\alpha)\widehat H_{\alpha\beta} \widehat
U_{\beta}^{-1}(\theta_\beta)\vert\Psi^\beta_{\theta_\beta}\rangle
= e^{-3i(\theta_\alpha + \theta_\beta)/2} \ \ &&\nonumber\\
\times\int
\Psi^\alpha_{\theta_\alpha}(re^{-i\theta_\alpha})H_{\alpha\beta}(r)
\Psi^\beta_{\theta_\beta}(re^{-i\theta_\beta}) r^2dr. \ \ \ \ &&
\label{1}
\end{eqnarray}
If the $\widehat H_{\alpha\beta}$ operator
connects rearrangement channels with the dynamical coordinates
$r_\alpha$ and $r_\beta$ [i.e.\ in the coordinate space
$\widehat H_{\alpha\beta}f(r_\beta)=\int dr_\beta r^2_\beta
H_{\alpha\beta}(r_\alpha,r_\beta)f(r_\beta)$], then
\begin{eqnarray}
\langle r_\alpha\vert\widehat U_{\alpha}(\theta_\alpha)\widehat
H_{\alpha\beta} \widehat
U_{\beta}^{-1}(\theta_\beta)\vert\Psi^\beta_{\theta_\beta}\rangle
=e^{3i(\theta_\alpha-\theta_\beta)/2}\nonumber \ \ \ \ \ \ &&\\
\times\int H_{\alpha\beta} (r_\alpha
e^{i\theta_\alpha},r_\beta)\Psi^\beta_{\theta_\beta}(r_\beta
e^{-i\theta_\beta}) r^2_\beta dr_\beta,\ \ \ &&
\end{eqnarray}
and
\begin{eqnarray}
\langle \Psi^\alpha_{\theta_\alpha}\vert \widehat
U_{\alpha}(\theta_\alpha)\widehat H_{\alpha\beta} \widehat
U_{\beta}^{-1}(\theta_\beta)\vert\Psi^\beta_{\theta_\beta}
\rangle =\ \ \
\ \ \ \ \ \ \ \ \ \ \ \ \ \ \ \ \ \ \ && \nonumber\\
e^{-3i(\theta_\alpha + \theta_\beta)/2}\int\int
\Psi^\alpha_{\theta_\alpha}(r_\alpha
e^{-i\theta_\alpha})H_{\alpha\beta}(r_\alpha,r_\beta)
\ \ &&\nonumber\\
\times\Psi^\beta_{\theta_\beta}(r_\beta e^{-i\theta_\beta})
r_\alpha^2dr_\alpha r^2_\beta dr_\beta. \ \ \ \ \ \ \ \ \ \ \
\ \ \ \ \ \ \ \ \ \ \ \ &&
\label{2}
\end{eqnarray}
Deriving (\ref{1}) and (\ref{2}) the Cauchy theorem was used,
assuming that all potential operators are analytic [like ours
(\ref{pot})]. We
can see that the matrix elements of the transformed operator
$\widehat U_{\alpha}(\theta_\alpha)\widehat H_{\alpha\beta}
\widehat U_{\beta}^{-1}(\theta_\beta)$ between the original
channel states $\vert \Psi^\alpha_{\theta_\alpha}\rangle$ and
$\vert \Psi^\beta_{\theta_\beta}\rangle$ can be expressed as the
matrix elements of
the original operator $\widehat H_{\alpha\beta}$ between the
so-called back-rotated channel states. This is a well-known
feature of the usual complex scaling method, too.

We tested this generalized CSM with the above two-channel
problem setting $\theta_2=0$. The result is in Fig.\
\ref{fig2}. The position of the $(-+)$ shadow pole remains the
same as was in Figs.\ \ref{fig1} within 7 decimal digits, which
is a  remarkable stability regarding that no optimization was
made in $\bar b$ and $\theta_1$.

The really relevant test is, however, an example where there is
a shadow pole on $(+-)$. To achieve
this, we set $\lambda_{22}=0.2$, which, in a one-channel
problem, results in a pole at $2.430-i3.704$ [Fig.\
\ref{fig3}(a)].
Switching on the  coupling ($\lambda_{12}=\lambda_{21}=0.42$),
figures similar to Fig.\ 1 can be generated [Figs.\
\ref{fig3}(b)--(d)]. Now the pole
at $3.381-i3.228$ is revealed when both channel continua have
swept over this point, so that this pole is on the $(--)$ sheet.
The other pole at $2.373-i3.357$ is revealed when the continuum
of the second channel has swept over it and that of the first
one has not, which shows that this pole is a shadow pole on the
$(+-)$ sheet,
in agreement with the fact that these poles originate from the
second channel \cite{Pearce}. In this example the variation of
the rotation
angles slightly removes the poles from their original positions.
This is, however, certainly caused by the fact that, because of
the unlucky location of the poles, we have to choose rotation
angles that are far from optimum. If we do not want to reveale
the two poles at the same time, we can optimize the
$\theta$ angles, which results in stable pole positions.

Finally, we mention an interesting feature of the present method.
Let us suppose that there is a multichannel problem where there
are degenerate thresholds. Then some of the Riemann
sheets cannot be reached from the physical sheet by following
analytical continuation paths because one cannot pass between
two thresholds that coincide. Using the above multichannel CSM
with different rotation angles in these channels, we can reach
such Riemann sheets.

In summary, we have investigated the applicability of the
multichannel complex scaling method to explore the pole
structure in multichannel scattering problems. We have
used a natural extension of the single-channel complex
scaling transformation to multichannel cases, which allows us to
find the poles of the scattering matrix on all Riemann
sheets. We have found that this extension works as expected
and is able to find all conventional poles and shadow poles
reliably.

This research was supported by OTKA (National Science Research
Foundation, Hungary) under contract No.\ 3010. The author is
indebted to Prof. B. Gyarmati and Prof. R.~G.
Lovas for stimulating discussions and for reading the
manuscript.

\figure{Energy eigenvalues of (a) a one-channel problem with
a one-term separable potential ($\lambda_{11}=1.0$); (b)--(d) a
two-channel problem with one-term separable potentials
($\lambda_{11}=1.0$, $\lambda_{22}=0.0$,
$\lambda_{12}=\lambda_{21}=1.0$, and $E_2$=2). The dots are the
points of the rotated discretized continua, while the circles
are the poles of
the $S$--matrix on different Riemann sheets. The rotation angles
(in radians) are: (a) 0.4, (b) 0.2, (c) 0.4, and (d) 0.7.
\label{fig1}}
\figure{Energy eigenvalues of the two-channel problem of
Fig.\ \ref{fig1}. The rotation angles are: $\theta_1=0.45$ and
$\theta_2=0$.\label{fig2}}
\figure{The same as Fig.\ \ref{fig1}, with the potential
strengths: (a) $\lambda_{22}=0.2$; (b)--(d) $\lambda_{11}=0.0$,
$\lambda_{22}=0.2$, and $\lambda_{12}=\lambda_{21}=0.42$. The
rotation angles are: (a) 0.78, (b) 0.68, (c) 0.78, and
(d) $\theta_1=0.42$, $\theta_2=0.78$.\label{fig3}}

\end{document}